\documentclass[manuscript]{aastex}

\newcommand{\lh}{\ensuremath{L_\mem{H}}}
\newcommand{\kelv}{\ensuremath{\,\rm K}}
\newcommand{\sak}{\mbox{V4334 Sgr}}

\newcommand{\lsun}{\ensuremath{\, {\rm L}_\odot}}
\newcommand{\teff}{\ensuremath{T_{\rm eff}}}
\newcommand{\msun}{\ensuremath{\, {\rm M}_\odot}}
\newcommand{\kpc}{\ensuremath{\, \mathrm{kpc}}} 
\newcommand{\mzams}{\ensuremath{M_{\rm ZAMS}}}
\newcommand{\hedr}{\ensuremath{^{3}\mem{He}}}

\newcommand{\czw}{\ensuremath{^{12}\mem{C}}}

\newcommand{\cdr}{\ensuremath{^{13}\mem{C}}}

\newcommand{\mem}[1]{\ensuremath{\mathrm{ #1}}}
\newcommand{\jahre}{\ensuremath{\, \mathrm{yr}}}
\newcommand{\etal}{et~al.\,}

\newcommand{\abb}[1]{Fig.\,\ref{#1}}

\slugcomment{Version: \today}

\usepackage{natbib}

\shorttitle{Sakurai´s evolution}
\shortauthors{F. Herwig}

\begin{document}

\title{The evolutionary time scale of Sakurai's object:\\ A test of convection theory?} 
\author{Falk Herwig
%\altaffilmark{1,2,3} 
}
\affil{University of Victoria, B.C., Box 3055, V8W 3P6, Canada}
\email{fherwig@uvastro.phys.uvic.ca}

\begin{abstract}
Sakurai's object (\sak) is a born again AGB
star following a \emph{very late thermal pulse}. So far no stellar
evolution models  have been able to explain the extremely fast
evolution of this star, which has taken it from the pre-white dwarf
stage to its current  
appearance as a giant within only a few years. A very high stellar
mass can be ruled out as the cause of the fast evolution. Instead the
evolution time scale is reproduced in stellar models by making the
assumption that the efficiency for 
element mixing in the He-flash convection  zone during the very late
thermal pulse is smaller than predicted by the mixing-length theory. 
As a result the main energy
generation from fast proton capture occurs closer to the surface and
the expansion to the giant 
state is accelerated to a few years. Assuming a mass of \sak\ of
0.604\msun\ -- which is 
consistent with a distance of 4\kpc --  a reduction of the mixing
length theory mixing  
efficiency by a factor of $\sim 100$ is required to match its
evolutionary time scale. This value decreases if \sak\ has a smaller
mass and accordingly a smaller distance. However, the effect does not
disappear for the smallest possible masses. 
These findings may present a semi-empirical constraint on the
element mixing in convective zones of the stellar interior. 

\end{abstract}


\section{Introduction}
Sakurai's object (\sak) has displayed a dramatically fast evolution
both in stellar parameters and in chemical abundance pattern
\citep{duerbeck:97,asplund:99a,duerbeck:00}. In 1976 it was possibly
detected by the  ESO/SERC survey close 
to the detection limit of m$_\mem{J}$=21 \citep{pollacco:99} 
which coincides with the
stellar parameters of a pre-white dwarf (WD) in the Hertzsprung-Russell
diagram (see \abb{fig1}). While this measurement is an important consistency check, 
the last non-detection in 1994 at the limiting magnitude of
m$_\mem{V}$=15.5 and the first positive detection at m$_\mem{V}$=12.4 by 
Takamizawa in 1995 \citep[see][]{duerbeck:97} represents a stringent
constraint on the  
evolutionary speed (\abb{fig1}). This evolution has been
interpreted as 
the result of a final He-flash which  occurred in 1994. By early
1996 the star had reached $\log 
L/\lsun \simeq 3.8$ and cooled to well below $\log \teff  \sim
4$. Since then it has continued to cool and brighten while displaying
RCrBr-like red declines. Thus, according to the observational
evidence, \sak\ must have 
completed the \emph{born again
  evolution} from the pre-WD stage to its
current appearance as a giant  in about two years. This interpretation
is supported by photoionization modeling of the planetary nebula of \sak\
\citep{pollacco:99,kerber:99}. These models place the central star at
an HRD location which is compatible with a pre-WD central star. 

The born again time scale of the related
object FG\,Sge has been successfully used to derive the stellar mass
under the assumption that the star has gone through a \emph{late
thermal pulse} 
\citep{bloecker:97}. Applying the same procedure to \sak\ leads to an
extremely large  
stellar mass of about 1\msun\ as noted by
\citet{duerbeck:00}. Such a high mass would require
the long distance scale ($d \simeq 8 \kpc$) in disagreement
with independent distance determinations \citep[such as the extinction
method;][]{kimeswenger:98}, which yield distances as low as
$d \simeq 1.1 \kpc$. 

There is more compelling evidence from nucleosynthesis that  \sak\ is not
very massive. The abundance ratio N/O 
in the planetary nebula (PN) is well below unity \citep{pollacco:99}. The PN
material reflects the envelope composition during the very last
phase on the AGB. According to recent stellar evolution
calculations by \citet{lattanzio:99}, AGB stars with the
highest mass have a continuously increasing N/O ratio due to hot-bottom
burning and eventually the ratio exceeds unity, e.g.\ at
$\mzams=6\msun$ in the case of solar metallicity. Therefore \sak\
cannot have the highest possible mass.

In fact, \sak\ must be of even lower mass
than required by the N/O constraint. \citet{asplund:99a} report a
significant lithium abundance  which  increased over
the six month period covered by their observation in 1996 ($0.5 - 1.0$
dex above initial solar). This amount of lithium cannot be inherited
from a previous evolutionary phase but must be a nucleosynthesis product
of the special conditions of the final flash \citep{herwig:00f}. Because
any mechanism of lithium creation relies on a readily available
reservoir of \hedr, the progenitor star must have avoided  hot-bottom
burning altogether and thus \sak\ is less massive than $\sim 0.7\msun$.

While observationally the very short evolutionary time scale with a
probably normal CSPNe (central star of PNe) seems well
established, a theoretical explanation of this phenomenon is
lacking. Here it should be noted that two different kinds of
models of the final flash have been constructed. Most 
models are  \emph{late thermal pulse} (LTP) evolutionary
sequences (which applies to FG\,Sge). The thermal pulse occurs
while the star is still on the horizontal crossing from the AGB to the
CSPNe phase at constant luminosity. During this first post-AGB phase
the H-shell is still active and prevents the mixing of envelope
material into the He-flash convection zone during the thermal pulse.
The born again evolution is energetically driven by the He-flash.
 For these LTP models $\tau_\mem{BA}$ (=  time of return to AGB - time
 of occurrence of LTP/VLTP) 
is about $100 - 200\jahre$ \citep{bloecker:95b,schoenberner:79} and
thus clearly in disagreement with the observation of \sak. 

Both the possible ESO/SERC detection in 1976 and the photoionization
models indicate that the post-AGB thermal pulse has occurred very late
when the star has already been approaching the WD cooling sequence
in the Hertzsprung-Russell diagram. In this case of a \emph{very late
  thermal pulse} (VLTP) the H-burning  has stopped and the protons in
the envelope are mixed down into the 
He-flash convection zone where they burn on the convective time scale.
Because the nuklear burning and convective mixing occurs
on the same time scale at the same location, a special
numerical treatment is required. Iben \etal (\citeyear{iben:83a}) 
circumvented this problem in their model calculation of the VLTP by
ignoring the nuclear energy released by proton captures
in the He-flash zone. Energetically their model resembles more that of the 
LTP and they found $\tau_\mem{BA}$ to be of the order of a few hundred
years (from their Fig.\,1), similar to the LTP born again
evolution. Another VLTP model sequence has been presented by Iben and
MacDonald (\citeyear{iben:95b}) with $\tau_\mem{BA} = 17\jahre$.
This is closer to the born again time scale of \sak, although
still too large by a factor of $3 - 4$. The difference
of the two values of $\tau_\mem{BA}$ is probably related to the
treatment of hydrogen burning.

The  VLTP sequence by \citet{herwig:99c} 
takes the nuclear energy generation by proton captures
in the He-flash convection zone into account. For this purpose a
numerical scheme has been developed which consistently couples the
nuclear network equations with the equations of time-dependent
convective element mixing \citep{herwig:00e}.
However, for this sequence we found 
$\tau_\mem{BA} \sim 350 \jahre$. Therefore, this model is not
in agreement with \sak. 

In this paper we demonstrate that the very short born again evolution time can
be reproduced by stellar models if the efficiency of element mixing
in the He-flash convection zone is reduced compared to the mixing
velocity predicted by  the mixing-length theory (MLT).

\section{Results}
\label{sec:results}

 The born again evolution of \sak\  is another case of the general
problem of why stars become red giants.
It is well known that the non-linear stellar structure 
equations as a boundary value problem have multiple solutions
which may be associated with different topologies \citep[e.g.][]{sugimoto:00}. 
If the assumption of thermal equilibrium is relaxed, the transition between 
solutions of different topologies can be obtained. This leads
to the initial value problem of stellar evolution. In order to
switch from a  dwarf structure to a giant structure the entropy of the
envelope has to be increased.  

In the VLTP model of \citet[][Fig.\,4]{herwig:99c} the
peak proton-capture energy release is located deep in the He-flash
convection zone. The entropy increase in these
layers by the additional 
H-burning luminosity barely affects the outermost layers because,
in the He-flash convection zone, the temperature is already greatly
increased by the ongoing He-flash. If the protons are captured at such a
deep position in the intershell region the corresponding energy release is
merely a perturbation of the prominent He-shell instability. Then, the
ingestion of protons does not significantly change the time scale of
the born again evolution. As a result the born again evolution
following a VLTP  will be -- rather similar to the born again
evolution following the LTP -- of the order of a few hundred
years. This seems not compatible with the observed time scale of the
born again evolution of \sak.

In order to construct born again stellar models with an evolutionary
speed in agreement with \sak\ we are looking for a modification of the 
stellar model which is capable of bringing the position of the main
H-energy release from ingested protons closer to the envelope. 
The position of peak hydrogen burning is determined by the competing
mixing ($\tau_\mem{mix}$) and nuclear time scales  ($\tau_\mem{nuc}$). 
In \citet{herwig:99c} we have adopted for the time dependent treatment 
of convective element mixing the diffusion coeffient
$D_\mem{MLT}=\frac{1}{3} \alpha_\mem{MLT} H_\mem{p} v_\mem{MLT}$ where
$v_\mem{MLT}$ is the convective mixing velocity according to the
MLT \citep{langer:85}. 
$\tau_\mem{nuc}$ decreases with increasing
temperature as the reaction rate of proton capture by \czw\ increases.
The main
energy generation by fast convective proton burning will occur at that
position 
in the He-flash convection zone where $\tau_\mem{nuc} \simeq
\tau_\mem{mix}$. This position moves
towards the top of the He-flash convection zone if the convective
efficiency of  mixing the isotopes is reduced. It can be expected that 
the 
speed of the born again evolution after a VLTP depends on the position
of H-burning energy release within the star.

In order to evaluate this hypothesis
new VLTP model sequences have been computed using  a starting
model of the original sequence  of \citet{herwig:99c} before the
ingestion of protons 
into the He-flash convection zone begins. The same evolutionary
code (EVOL) has been used. The time evolution of 16
isotopes for hydrogen- and  
helium-burning is followed. Time-dependent overshoot on any convective
boundary 
can be considered and the latest OPAL opacities have been used
\citep{iglesias:96}. The mixing length parameter is
$\alpha_\mem{MLT}=1.7$ and the  metallicity is $Z=0.02$.

One way to reduce the convective mixing efficiency would be to change
the mixing length parameter $\alpha_\mem{MLT}$. However, this 
affects both the convective transport of energy as well as that of
matter. According to our hypothesis we are only interested in the
efficiency of material transport. Besides, the mixing length parameter 
describing the efficiency of convective energy transport
has been calibrated in order to reproduce the solar parameters.
It does not seem plausible to change the well established 
parameter $\alpha_\mem{MLT}$. Contrary, the effect of the efficiency of 
mixing material is much less obvious in stellar evolution
calculations. In most situations the nuclear time scale is larger than 
the convective mixing time scale by orders of magnitude. E.g.\ for main
sequence stellar models the velocity of convective material transport
can be changed by considerable factors without any change to the
stellar parameters. 
Therefore we define a new parameter $f_\mem{v} \equiv D_\mem{MLT} /
D_\mem{CM}$ where 
$D_\mem{CM}$ is the diffusion coefficient for composition mixing.
VLTP and born again model sequences for $f_\mem{v}=1$, $3$, $30$ 
and $300$ have been computed. A 0.535\msun\ VLTP sequence 
with a \mzams=1\msun\ progenitor model \citep{herwig:00c}  and $f_\mem{v}=1$ 
has been computed for comparison.  

The computations show that 
the evolution across the HRD is accelerated for larger values of
$f_\mem{v}$. The cases $f_\mem{v}=3$   
and $f_\mem{v}=30$ for the mass 0.604\msun\ are shown in \abb{fig1}.
The AGB is reached within $195$ and $9.77\jahre$ respectively. 
In a VLTP model sequence with  $f_\mem{v}=30$ the peak p-capture
energy is released at $m_\mem{r}\sim 0.601\msun$ compared to 
$m_\mem{r}\sim 0.595\msun$ with $f_\mem{v}=1$.

Models with reduced convective velocity for composition mixing do not
only evolve faster, but they also feature a modified evolution of 
convective zones. In the 
new computation with the reduced convective 
mixing efficiency, H-burning takes place in the top layers of the
intershell convection zone and establishes its own convective layer on
top of the actual He-flash convection zone. In contrast H-burning in
the original computation takes place deeper inside the intershell region and
the separate H-burning convection zone inside the He-flash convection
zone is very short lived. 

A quantitative comparison of the evolutionary times for different
$f_\mem{v}$ values with observed evolution times is given in \abb{fig3}. 
Observationally two time intervalls can be defined: from the last
non-detection in 1994 to the first positive pre-discovery  
detection in 1995 (t2) and from this time to the date of the first spectra 
reported by \citet{asplund:99a} (t1). If \sak\ has a mass of 0.604\msun\ then 
this tentative comparison requires a reduction of the efficiency of
composition mixing   
 of $f_\mem{v} \sim 100$. This follows from the comparison of time 
interval t1. The fact that time interval t2 requires a larger reduction factor 
is due to the fact that the observed time interval t2 is only a
lower limit because  \sak\ was at or below the detection limit in 1994.

\section{Discussion}
\label{sec:concl}
Several tests have been carried out to ensure obvious sources of uncertainty
do not jeopardise the general validity of the  findings. 
Qualitatively, the results are independent on the choice of opacities
and the assumptions on overshooting.
The influence of numerical parameters, such as the mass at
which the outer and the interior solutions are attached,
do not change the results qualitatively. 

However, several improvements are necessary in future models. 
We have not considered the $\mu$-barrier during the ingestion of
protons which might effect the results to some
extent. Moreover, time-dependent treatment of convective energy
transport is not considered. This means that the
time step has always to be chosen sufficiently larger than the
convective mixing time scale according to the MLT in order to remain
consistent with the assumption of instantaneous convective energy
transport. This criterion prohibits the usual time resolution which,
for instance, requires that the hydrogen-burning luminosity
$L_\mem{H}$ may not increase by more than a few percent. In the VLTP
case $L_\mem{H}$ often multiplies by some factor within one time
step. However,  the nuclear energy integrated by the
structure equations and the nuclear energy estimated from the amount
of consumed proton agree within 10-20$\%$
in all cases. 

The reduced efficiency of composition mixing leads to a different evolution
of the convective zones.
This affects 
the CNO element and isotopic ratios. The surface $\czw/\cdr$ ratio of the 
$f_\mem{v}=30$ born again model is $\sim 5$ and thus in agreement with
the observations of \citet{asplund:99a}. Also the
CNO elemental ratios of this sequence are in good agreement with the
observed ratios. The absolute
CNO abundances of \citet{asplund:99a} are smaller by a factor of
$\sim5$ compared to the model predictions. The reason for this
inconsistency is not clear and might be due
to the so-called carbon problem of abundance analysis \citep{asplund:00}.

The formation of lithium is another important test for any stellar model
sequence of Sakurai's object. More detailed studies of the 
convective nucleosynthesis  should provide additional
constraints on the validity of the proposed concept.
From a prelimary analysis of the temperature conditions in models with
reduced convective element mixing we expect that the mechanism 
of hot hydrogen-deficient \hedr-burning \citep{herwig:00f} for the 
synthesis of lithium during the VLTP will provide  a lithium
abundance in agreement with that of \sak.

It has been shown that the born again
evolution following a VLTP  
is sensitively dependent on the composition mixing due to convection.
Therefore, it is important to consider any mechanism or parameter on 
which the convective velocity in this region depends. 
The computations show that the convective velocity in the He-flash convection 
zone decreases with stellar mass of the post-AGB star. While the
0.604\msun\ displays $v_\mem{MLT}\sim 3 \mem{km/s}$ (at $\Delta m=0.003\msun$ 
below the top boundary of the He-flash convection zone) the comparison model 
sequence of mass 0.535\msun\ shows only $v_\mem{MLT}\sim 0.35 \mem{km/s}$.
In accordance with the previous finding the 0.535\msun\ model sequence does 
evolve much faster back to the AGB than the 0.604\msun\ sequence (\abb{fig2}).
This faster evolution must be attributed to the lower convective velocity.
The evolutionary times for the previously described intervals t1 and t2 are 
included in \abb{fig3} and lie clearly off the relation between $f_\mem{v}$ 
and $t_\mem{BA}$ for 0.604\msun. However, note that $v_\mem{MLT}$ of the 
0.535\msun\ sequence is just about a factor of 10 smaller than in the
0.604\msun\ case. Possibly all these models follow a narrow relation 
between $t_\mem{BA}$ and $v_\mem{CM}$. 

In any case, the assumption of a small mass for \sak\  probably
 does not solve the 
time scale problem alone. The time scale for this 
model sequence is still too large by an order of magnitude. Moreover, stellar
masses as low as 0.535\msun\ are not consistent with the possible
detection in the ESO/SERC in 1976 (\abb{fig2}). 
However, this argument depends somewhat on the distance-luminosity
relation used. Finally, the 0.535\msun\ evolutionary track is not
consistent with the photoionization models of
\citet{pollacco:99} and \citet{kerber:99}. 
Note, that the horizontal luminosity of the 0.604\msun\ post-AGB track
is somewhat larger than that of previous computations of comparable
mass because of the inclusion of 
AGB overshooting and the resulting effects on the core-mass luminosity
relation  
described by \citet{herwig:98b}. 

\section{Conclusion}
We have constructed a new born again stellar
evolution model 
for \sak\ which can reproduce the observed evolutionary speed across
the HRD as well as the CNO abundance ratios. It is probably compatible 
with the observed lithium abundance. For the new model it has been
\emph{assumed} that the efficiency of material transport by convection is
smaller than predicted by the MLT by a factor of $\sim100$ (depending
on mass). About the underlying physical origin of such a reduction we
can only speculate. The results presented here suggest that a
modification to the prescription of convective element mixing may be 
a solution for the otherwise incomprehensible
evolutionary behaviour of \sak.

\acknowledgments
I would like to thank D.A.\,VandenBerg for very useful
discussions and for his encouragement of this work. 
 Support through his Operating Grant from the Natural
Sciences and Engineering Research Council of Canada is gratefully acknowledged.
Moreover, it is a
pleasure to thank
T.~Bl\"ocker, W.-R.\ Hamann, L.\ Koesterke, N.\ Langer, D.\ Sch\"onberner and M.\ Steffen for the
very enjoyable collaboration on related projects which has paved the
way for this paper. R.\ Napiwotzki has
provided bolometric corrections of his model atmospheres of CSPN.

%\bibliography{astro}

\clearpage

\begin{figure}
\plotone{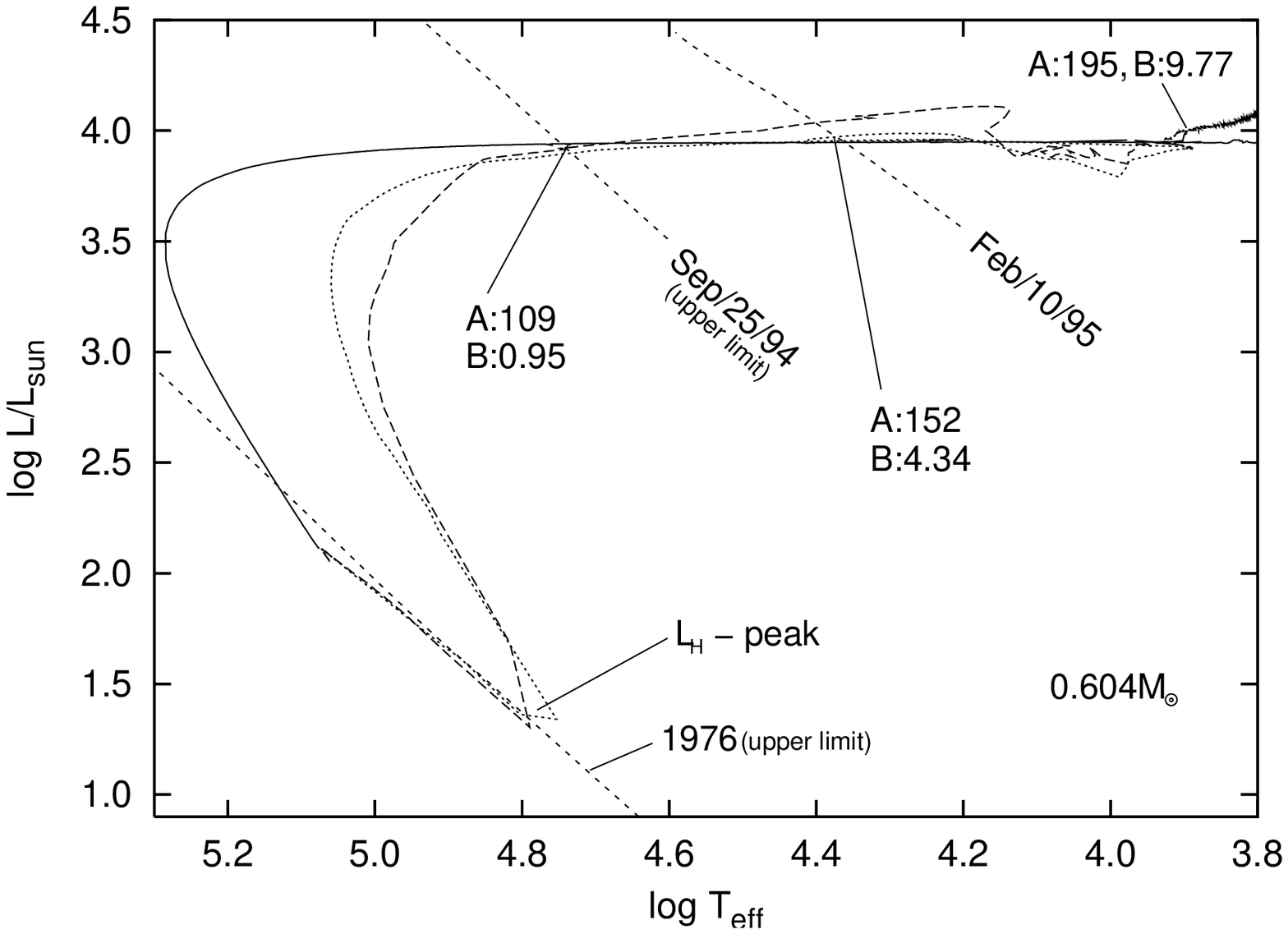} %--> HRD-VLTP-0.604.eps
\figcaption{ Hertzsprung-Russell diagram of the post-AGB
      sequence of mass $0.604\msun$ of Herwig \etal (1999) (solid line)
      with the VLTP  and 
      subsequent born again evolution 
      recomputed with $f_\mem{v}=3$ (dotted line, time labels
\textsf{A:}) and $f_\mem{v}=30$ 
      (long-dashed, time labels \textsf{B:}). 
      Time labels given at $\log \teff = 3.9$
      correspond to the first spectra presented in
      \citet{asplund:99a} which are dated Apr 20, 1996. All time
      labels are in \jahre\ and give the evolutionary time from the
      moment of largest H-burning luminosity \lh-peak. The  dashed
      straight diagonal lines represent three different magnitudes at
      or below which \sak\ has been observed at different times (see
      text). In this figure $d=4\kpc$ has been assumed for internal
consistency with the luminosity of the 0.604\msun\ evolutionary track.
The distance has been
      determined from the luminosity-distance relation derived from the data
      of \citet{duerbeck:00} for the date of the first spectra found in 
      \citet{asplund:99a} and the corresponding luminosity of the evolutionary
      tracks at the temperature determined from the spectra. For the lines of
      constant magnitudes the following has been used: for $\teff >
      40000\kelv$ B.C.$_\mem{V}$ from 
      \citet{napiwotzki:01} and for $\teff <
      40000\kelv$ indices from \citet{bessel:98} (Kurucz atmosphere
      models), (V-J)=-0.75, $A_\mem{J}=0.29A_\mem{V}$,
      $E_\mem{B-V}=0.7$. Two lines are labeled \textsf{upper limit} because \sak\ has been below or close to the detection limit.  \label{fig1}}   
\end{figure}

\begin{figure}
\plotone{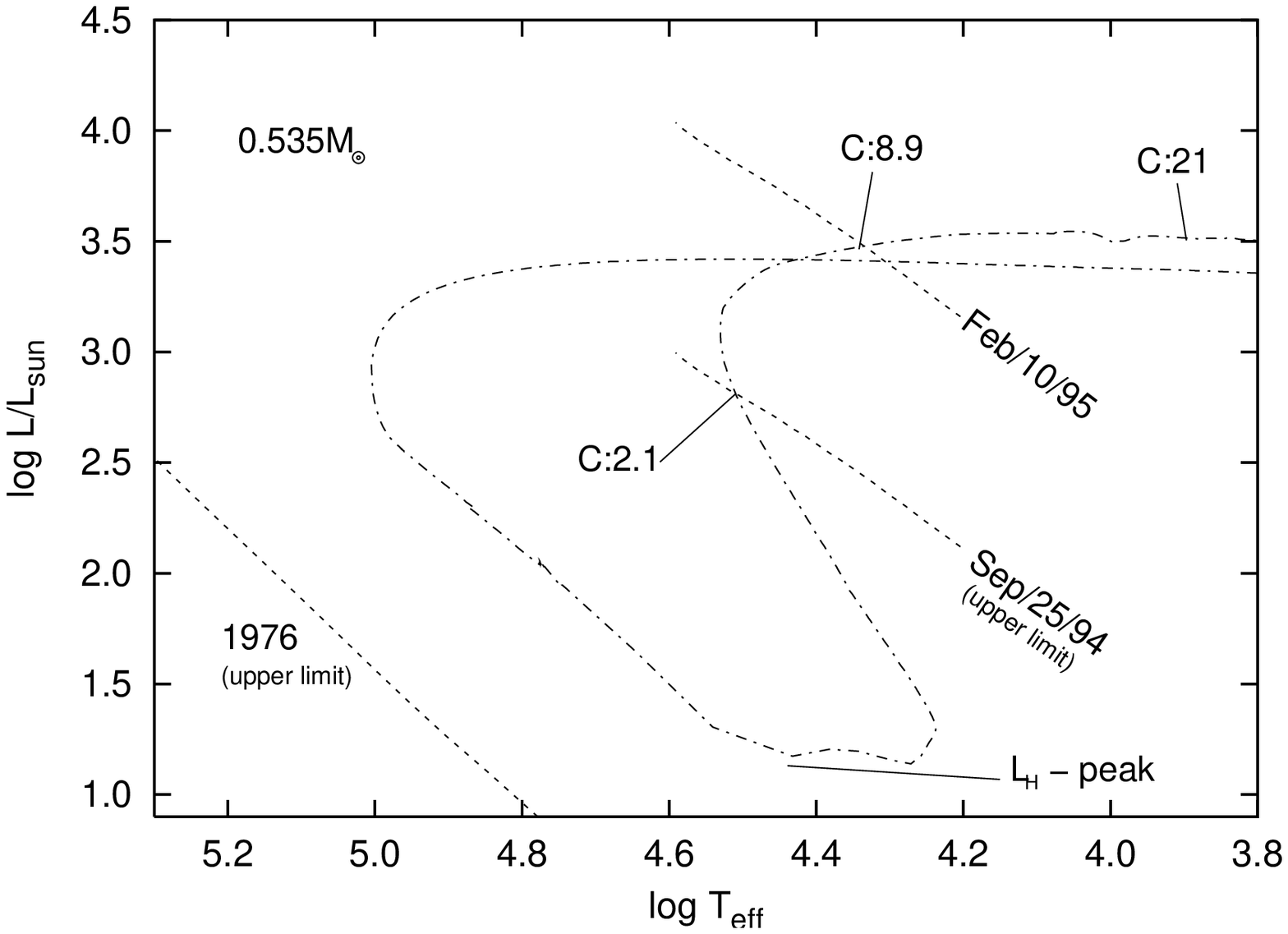} %-->  HRD-VLTP-0.535.eps
\figcaption{ The same as \abb{fig1} for  a $0.535\msun$ post-AGB evolution with
      VLTP ($f_\mem{v}=1$, time labels \textsf{C:}). 
      The  dashed
      straight diagonal lines again represent different magnitudes at
      or below which \sak\ has been observed. For this mass a distance
of $d=2.5\kpc$ is consistent with the luminosity of the 0.535\msun\
track. \label{fig2}}   
\end{figure}
\begin{figure}
\plotone{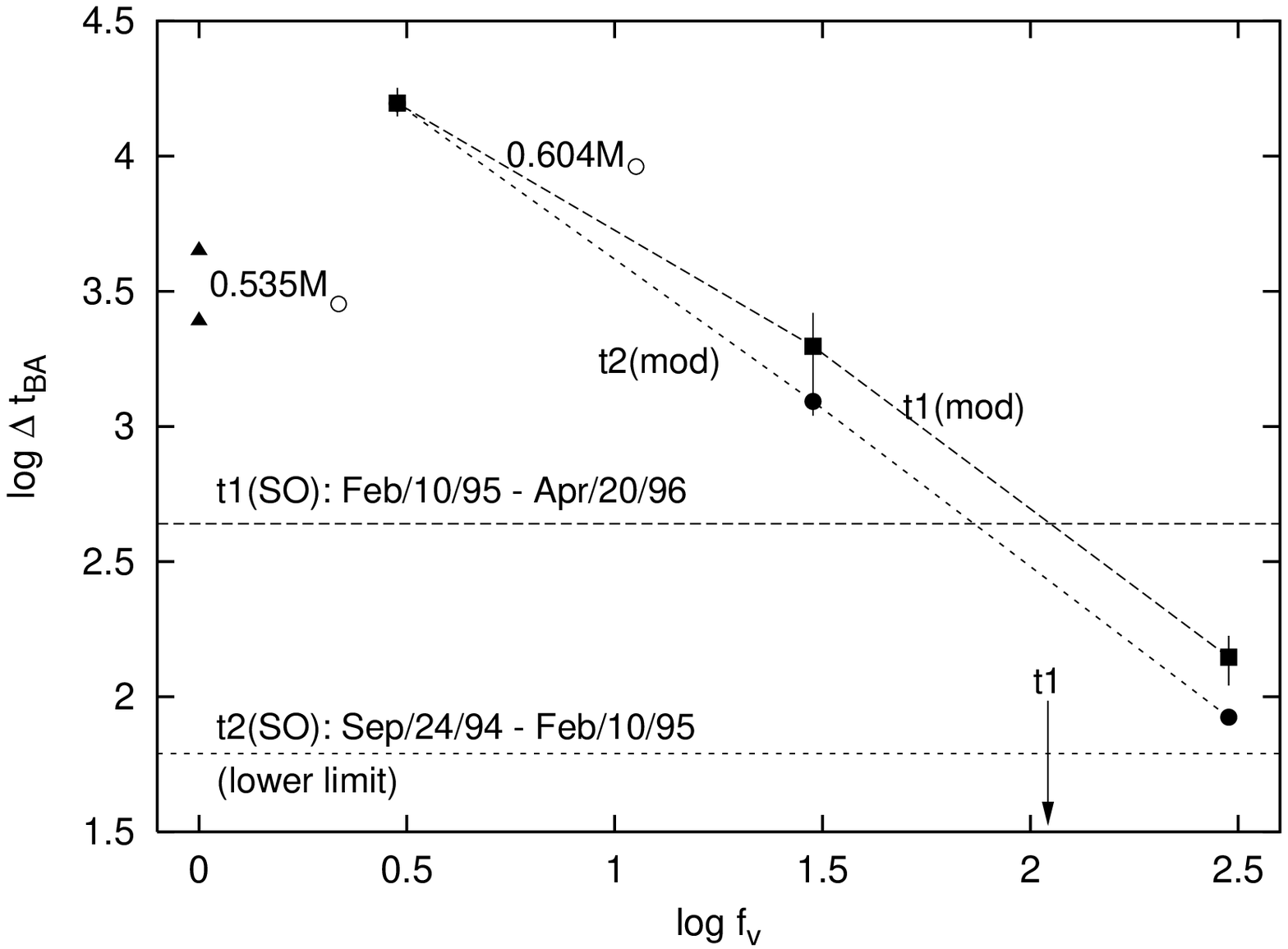} %--> f_v-PLOT/f_v-t_BA.eps
\figcaption{
Comparison of observed evolutionary speed of \sak\ and
 of born again evolution models. The horizontal lines labeled
  t1(SO) and t2(SO) mark the length of two time intervalls between
  observations of \sak. The dates of the observations are given beside the
  lines (compare \abb{fig1}).    The line t2(SO) is labeled
  \textsf{lower limit} because the corresponding line (Sep/24/94) of constant
  magnitude in \abb{fig1} is an upper limit for \sak\ at that date.
  From the 0.604\msun\ model sequences with different values
   of $f_\mem{v}$ the corresponding time intervals have been extracted
  (filled symbols)  and connected by lines labeled \textsf{t1(mod)} and
  \textsf{t2(mod)}. These lines show that
  born again models with larger reduction factors evolve faster between to
  given locations in the HRD.
  Error bars are plotted for squares only
  and reflect an estimate of observational uncertainties.
  The triangulars represent the evolutionary times of the 0.535\msun\
  sequence (top triangular belongs to interval t1). The comparison of the model
  data and the observed data yields $f_\mem{v}\sim 100$ if \sak\ has a
mass of 0.604\msun.
\label{fig3}}
\end{figure}

\end{document}